\documentclass[debug]{rmaa}
\usepackage{paralist}

\usepackage{psfrag,color}

\usepackage[latin1]{inputenc}
\usepackage{epstopdf}
\title{Spectral Properties and Variability  of BIS objects.\altaffilmark{1} } 

\author{
S. Gaudenzi,\altaffilmark{2} 
R. Nesci,\altaffilmark{2} 
C. Rossi,\altaffilmark{3,4}
S. Sclavi,\altaffilmark{4} 
K. S. Gigoyan,\altaffilmark{5} 
and  A. M. Mickaelian,\altaffilmark{5} 
  }

\altaffiltext{1}{Observations collected
 with the Cassini Telescope at Loiano station of the 
 INAF$-$Bologna Astronomical Observatory, 
 with the Copernico Telescope of the INAF$-$Padova Astronomical Observatory 
 and with the TACOR Telescope of the Universit\`a ``La Sapienza'', Roma.
}
\altaffiltext{2}{INAF/IAPS, Roma Italy.}
\altaffiltext{3}{INAF/ Osservatorio di Monte Porzio, Roma, Italy.}
\altaffiltext{4}{Universit\`a La Sapienza Roma, Italy.}
\altaffiltext{5}{ Ambartsumian Byurakan Astrophysical Observatory (BAO). }

\shortauthor{ Gaudenzi, Nesci \& al.}
\shorttitle{ BIS sources : optical results.}

\fulladdresses{
\item S.Gaudenzi and R.Nesci: INAF/IAPS, via Fosso del Cavaliere 100, 00133 Roma, Italy  (silvia.gaudenzi@iaps.inaf.it); (roberto.nesci@iaps.inaf.it).
\item C. Rossi: INAF-Osservatorio Astronomico di Roma, Via Frascati 33, 00040, Monte Porzio Catone (RM), Italy (corinne.rossi@uniroma1.it).
\item S. Sclavi and C. Rossi: Dipartimento di Fisica, Universit\`a La Sapienza,
Piazzale Aldo Moro 3, 00185 Roma, Italy.
\item K.S.Gigoyan and A.M. Mickaelian: V. A. Ambartsumian Byurakan Astrophysical Observatory (BAO) and Isaac Newton Institute of Chile, Armenian Branch, Byurakan 0213, Aragatzotn province, Armenia.
  }

\listofauthors{ S.Gaudenzi, R.Nesci  \& al.}
\indexauthor{Gaudenzi, S.}
\indexauthor{Nesci, R.}
\indexauthor{Rossi, C.}
\indexauthor{Sclavi, S.}
\indexauthor{K. S. Gigoyan.}
\indexauthor{Mickaelian, A.M.}

\abstract{
Through the analysis and interpretation of newly obtained and of literature data we have clarified the nature of poorly investigated $IRAS$ ~point sources classified as late type stars, belonging to the Byurakan $IRAS$ ~Stars catalog.  From medium resolution spectroscopy
of  95 stars   we have strongly revised 47 spectral types and  newly classified 31 sources.
Nine  stars are of G or K types, four  are N carbon stars in the Asymptotic Giant Branch, the others  being M$-$type stars.
 From literature  and new photometric observations we have studied their variability behaviour.
 For the regular variables we determined distances, absolute magnitudes and mass loss rates. For the other stars we estimated the distances, ranging between 1.3 and 10~kpc with a median   of 2.8~kpc  from the galactic plane, indicating that BIS stars mostly belong to the halo population.    
   }

\resumen{
 Un gran número de entradas del catálogo de fuentes puntuales $IRAS$ tienen todavía una clasificación aproximada.
A través del análisis y la interpretación de nuevos datos obtenidos y datos de la literatura tuvimos como objetivo clarificar la naturaleza de las pobremente investigadas fuentes puntuales de $IRAS$ clasificadas como estrellas de tipo tardío, pertenecientes al catálogo de estrellas Byurakan $IRAS$.
En 95 estrellas realizamos espectroscopía de mediana resolución para mejorar la clasificación espectral que previamente se había basado solo en índices de color en el óptico y/o espectros de baja resolución. Recolectamos datos fotométricos de la literatura y obtuvimos nuevas observaciones fotométricas para estudiar su variabilidad; asignamos o revisamos la clasificación de variabilidad de nuestros objetos.
Revisamos los tipos espectrales y encontramos diferencias sustanciales con respecto al catálogo BIS original en 45\% de las estrellas; además el 30\% no tenían clasificación previa.
Para las estrellas variables regulares determinamos sus distancias, sus magnitudes absolutas y sus tasas de pérdida de masa. De las otras estrellas, estimamos sus distancias a partir de sus tipos espectrales y sus magnitudes aparentes: Se encontraron distancias desde 1.3 a 10 ~kpc, con una distancia media desde el plano Galáctico de 2.8 ~kpc, lo cual indica que las estrellas del catálogo BIS en su mayoría, pertenecen a la población del halo de la Galaxia. Un gran número de estas estrellas se encontrarían al alcance de una medición directa de GAIA.
Los espectros muestran que nueve estrellas son de tipo espectral G-K, 
cuatro estrellas de Carbón N se encuentran en la rama gigante asintótica, las otras  son del tipo espectral M.
}

\addkeyword{}
\addkeyword{Stars: late type}
\addkeyword{Stars: variables}
\addkeyword{Stars: Mass loss}

\begin{document}
% Typeset article header
\maketitle

\section{Introduction: Byurakan -- $IRAS$ ~Stars catalog}  \label{Intro}

Asymptotic Giant Branch (AGB) stars are very luminous objects, in a fast evolutionary phase which mix up the atomic nuclei produced by nuclear reaction in the inner part of the star into the external envelope and then into the interstellar space through the stellar wind and the final formation of a Planetary Nebula.
A better census of the AGB stars population and good knowledge of their physical characteristics and variability is a basic requirement to perform satisfactory checks of our models of stellar evolution and to study the chemical evolution of our Galaxy \citep{Lebz12, Bladh13, Now13}. The forthcoming direct distance measures by the GAIA mission of the stars within several kpc from the Sun will allow a substantial improvement in this field.
With this aim in mind, we performed a study of a sample of AGB candidates collected in the Byurakan Infrared Stars (BIS) catalog \citep{mic06}, based on the spectra of galactic sources visible in the objective prism plates of the First Byurakan Survey (FBS)  \citep{mark89} and on the $IRAS$ ~Point Source Catalog \citet{PSC88}.

In 1983, the Infra Red Astronomical Satellite ($IRAS$) surveyed about 96\% of the 
sky in bands centred  at 12, 25, 60, and 100$\mu$m. 
More than 245000 point sources were detected and their fluxes and positions 
were listed in the $IRAS$ ~PSC. A recent work by \citet{Abr15} made a cross-check  of the {$IRAS$ Point Source Catalog (PSC) and Faint Sources Catalog (FSC) \citet{Mos89}  improving their positions and correlation with infrared sources detected by more recent missions ($WISE$ \citep{Cut12} , $AKARI$ \citep{Ish10}, 2MASS \citep{Cut03}).
The $IRAS$ ~archive also  includes several spectra in the Low$-$Resolution Spectra catalog \citep{LRS87} in the range 7.7$-$22.6~$\mu$m  providing useful indications on the chemical composition of the dust shells around the stars (see \citealp{olnon86}, and the \citealp{IRASEXP}).

The   BIS catalog  contains data for a final census of  276 IR sources 
as being potentially stars of late spectral types. 
 The stars were selected on the basis of their low resolution spectra of FBS and of the Dearbon Astronomical Observatory \citep{lee47}.
Images  of the Palomar Observatory Sky Survey (DSS,  \citealp{Abe59} ) were also used to check the identification.  The most recent citations of literature spectral types are to be found in \citet{skiff16}.   
 
 The identifications on the FBS plates were carried out in the region with $\delta$ $>$ +61$^o$ and galactic latitude b $>$ +15\arcdeg,  covering  a surface of 1504~deg$^2$.  

For each object several information are given, such as accurate optical positions for two epochs (B1950 and J2000), photometric data after cross$-$correlation with MAPS \citep{Cab03}, USNO$-$B1.0 \citep{monet03} and 2MASS catalogs, proper motions (PM) and a  classification based on the Digitized First Byurakan Survey spectra (DFBS, \citealp{mic07}) accessible online from the webpage http://ia2.oats.inaf.it/ . The stars have intermediate values of galactic coordinates, ranging in longitude between 90 and 151 degrees, and in latitude between 14 and 45 degrees, with average value of 30 degrees: they are therefore outside the solar circle and  are likely members of   the halo / thick$-$disk population.

Aiming at  clarifying the nature of the stars included in the BIS catalog we made a systematic collection of all the information available in literature and performed the acquisition of new photometric and spectroscopic data.

This paper is devoted to the optical properties of a subsample of 95 stars randomly selected, while a companion  paper will be dedicated to the infrared characteristics  of all the stars of the  BIS catalog  using data collected from public archives. 
\S~\ref{OLTS}, describes our targets; 
\S~\ref{Obs}, and \ref{DA}, are devoted to the observations and data analysis;
\S~\ref{PP}, is devoted to estimate and discussion of  the parameters of the IR sources;  
our conclusions are summarised in \S~\ref{Conc}. 

\section{Our targets}  \label{OLTS}

The  BIS catalog  contains several types of sources, including  M and CH stars, Mira$-$type and Semi$-$Regular (SR) variables, OH and SiO sources, N$-$type carbon stars and unknown sources surrounded by thick circumstellar shells.
 Most of our targets are poorly studied both from the photometric and spectroscopic point of view and only a few of them are classified as variables in the General Catalog of Variable stars (GCVS, \citealp{Sam17}) or in the Variable Star Index (VSX, \citealp{watson16}) catalogs. A preliminary clue of  variability  was given in  \citet{mic06} using the differences B1$-$B2 and R1$-$R2 of the two epochs of the DSS as reported in the USNO-B1.0 catalog.
 
 Among the 276 stars of the BIS catalog, 13 have color index 1.5$\le$ B$-$R$ \le$2.5,
 the remaining have B$-$R $\ge$ 2.5~mag.
We obtained new  optical CCD multi$-$band photometry and medium resolution spectra of a  sample  of  95 stars:
we have randomly selected 9 with color index $ \le$2.5 in order to check the  reliability of the original classification and to classify two stars  having  no previous classification.
We also obtained the spectrum of BIS028, already known to be a planetary nebula. 

\section{Observations}    \label{Obs}

 Spectra  in the range 3940-8500~\AA, 3.9~\AA/pixel dispersion, were obtained with the Cassini telescope of the Bologna Astronomical Observatory (Italy) at Loiano, equipped with the Bologna Faint Objects Spectrometer and Camera (BFOSC) and a EEV P129915 CCD. Photometric observations were also obtained with BFOSC, possibly on the same dates as for spectra. Some objects were observed also at the Copernico Cima Ekar telescope of the Padova Astronomical Observatory (Italy) equipped with the Asiago Faint Objects Spectrometer and Camera (AFOSC) mounting a  TK1024AB  CCD. All the stars were always observed in the Red Johnson filter.

Photometric observations were also performed in the $R$ filter in the period July$-$November 2011 with the TACOR~\footnote{TACOR = Telescopio A COntrollo Remoto} telescope of the Department of Physics of the University ``La Sapienza'' in Rome equipped with an Apogee U2  CCD. The data were reduced by means of standard IRAF procedures~\footnote{IRAF is distributed by the NOAO, which is operated by AURA, under contract with NSF.}.

Table~\ref{tab:tab1} presents the journal of observations; the columns have the 
following meaning: 
 1~- BIS number in the catalog;
 2~- $IRAS$ ~FSC designation of the objects;
 3~- date of observations;
 4~- spectral type according to the classification from our CCD spectra;
 5~- Red magnitude at the date of observation and observatory:
 ~$^1$ Loiano; ~$^2$ Cima Ekar; 
all the other data were obtained with the TACOR telescope.
 6~- magnitude range from the archive of the Northern Sky Variability Survey 
     (NSVS, \citealp{woz04a}; see Section~\ref{PV}); 
 7~- our variability classification defined in Section~\ref{PV}.  

\smallskip

A large number of spectroscopic standards (from K7 to M9 giants and dwarfs) were observed with the same instrumental configuration as our target stars and used as the basis for our classification. We have selected the templates of M standards from several catalogs and Spectral Libraries 
%\citet{luy79}; 
% \citet{GunSty83}; 
\citep{kirk91, torres93, fluks94, allen95, gray09}. 
Spectra of M standards were also downloaded from the site  http:$//$kellecruz.com$/$M\_standards$/$.  For the carbon stars we used \citet{barn96} and \citet{totten98}.

\section{Data analysis}   \label{DA}

%In addition to our new data, we collected all the known data available in literature.
In the following sections we describe the general characteristics of the spectra and the optical light curves for individual stars, when available.
 
\subsection{General characteristics for spectral classification} \label{GCfSC}

 Spectral types were derived by overlapping  the spectral tracing 
of the targets with those of the reference stars, overplotting them with IRAF/splot. Classification was made independently by three of us and the typical uncertainty is one subtype. We have checked the self-consistency of the reference stars in the same way. 
 Most of  the reference stars are  nearby. The  spectral resolution used   allows to discriminate luminosity classes  but not the metal content.

The spectra of almost all our targets are typical for M$-$type stars. In these stars the most prominent absorptions belong to the \ion{TiO}\,bands at 4761, 4954, 5167, 5448, 5862, 6159, 6700, 7055 and 7600~\AA. 
In some cases the \ion{VO}\,bands of the red system are also present, with several band heads in the range 7334$-$7472~\AA, and 7851$-$7973~\AA, seen only in very late type stars.
We paid special attention in looking for features used as typical dwarf/giant discriminators, like the Mg $b$ triplet 5167, 5173, 5184~\AA; the NaD doublet 5890, 5896~\AA; the \ion{CaOH}\, diffuse bands centered at 5550 and 6230~\AA, the \ion{MgH}\,bands at 4780, 5211~\AA; the \ion{CaH}\,at 6382, 6908, 6946~\AA \ \citep{mould76, giov94}. Few atomic lines belonging to Fe and Ti are also present in some spectra. 

From their spectral characteristics  four targets appears to be N$-$type AGB carbon stars. Three of them are embedded in a dense envelope with the blue region strongly  underexposed.

Two stars, BIS\,104  and BIS\,106, were erroneously classified as carbon stars on the basis of the $IRAS$ spectrum \citep{guglielmo97} while our spectra clearly show the typical features of intermediate M-type stars.

The nine stars in our sample  with  B$-$R$ \le$2.5  showed spectra typical for G and early K type stars, as expected. 
  BIS\,094, 105, 109, 229,  251, 259, and 286 are  in fair agreement with the old classification;
 BIS\,060 and BIS\,131 had no previous classification.   BIS\,131, has an infrared excess and has been the subject of a previous publication \citep{Ros10}. All these stars  will not be discussed further.  

None of our stars appears to be of luminosity class V (Main Sequence), one is  of class I and two of class II.

Our spectra are by far of too low resolution to measure the radial velocities of stars, so we cannot give a kinematic indication of the kind of stellar population (halo, thick disk) to which our stars belong.
None of the stars have an appreciable ($\ge$20mas) proper motion in the USNO-B1 catalog. 

 Figures  \ref{Fig1} and~\ref{Fig2} show a selection of  representative  M type spectra, the other spectra of the same class being similar to those  presented.
The atmospheric absorption bands of \ion{O$_2$}\,at 6867 and 7594~\AA, and of 
\ion{H$_2$O}\,at 7186~\AA, are not removed. In these figures  the ordinates are relative intensities corrected for atmospheric extinction, normalised with the maximum set to 100.

\subsection{Photometric variability}  \label{PV}

Most of our targets are present in the NSVS database, collected between 1997 and 2001 from the 
ROTSE$-$I (Robotic Optical Transient Search Experiment I) experiment \citep{woz04a}.
 The  observations  spanning up to one year gave us reliable indications about the photometric variability of our sample.
We have downloaded all the available light curves from the NSVS web site 
%\footnote{http://skydot.lanl.gov/nsvs/nsvs.pht/ } 
to check the photometric behaviour and compare the ROTSE magnitudes (R$_{r}$) with our new data. 
Some objects are not present in the NSVS archive,  being fainter than ROTSE$-$I detection limit (15.5~mag). 
 Few  stars have been observed by the  Catalina Real$-$time Transient Survey
(CRTS~\footnote{http://www.lpl.arizona.edu/css/}).  
Magnitudes from other catalogs  have been only considered as 
indicative, being obtained from the  DSS plates where many  stars are saturated.

It is worth to remember that the NSVS data were obtained with an unfiltered CCD, so that the quantum efficiency of the sensor makes the effective band most comparable to the Johnson $R$ band \citep{woz04a}, or better a mix of $V$ and $R$ colors, which is a function of the spectral type of the star. To inter$-$calibrate the NSVS and our magnitudes we used the stars in our sample with a very stable NSVS light curve. Our photometry is tied to the $R$ magnitude scale of the GSC2.3.2 catalog \citep{Lask08}, and a good calibration would require a bigger set of non variable stars in the M0$-$M8 spectral types range to define a reliable color correction. 
In fact, we have verified that for M8 stars, which emit most the photons in the IR tail of the unfiltered detector, the ROTSE instrumental magnitude are generally brighter than for M1 stars of similar $R$ magnitude. In any case, even with this {\it caveat}, our data have been useful to confirm the 
variability/stability of the stars of our sample. 

We assign  three main variability indices on the basis of the light curves: \\
{\bf (1)}{\rm ~regular, large amplitude variables, larger than 1.1~mag};\\
{\bf (2)}{\rm ~irregular, large amplitude variables, between 0.5 and 1.1 mag}; \\
{\bf (3)}{\rm ~ small amplitude  variables stars ( up to 0.5~mag) or
non$-$variable ( up to 0.3 mag)}.\\
  A few stars  show small amplitude either irregular or quasi regular  variability. To these stars we assigned intermediate  variability classes (2/3, 2/1), reported in Table~\ref{tab:tab1}  only. 
 %The light curves  are available at the NSVS web site.
 Here below we grouped stars with similar characteristics to avoid useless repetitions. For a number of stars we add spectroscopic and/or   photometric details.

%
%   FIGURA  Emission line stars
%
%\setcounter{figure}{7}
\begin{figure*}%  \includegraphics[width=\columnwidth]{spMemi.eps}
    \includegraphics[width=\columnwidth]{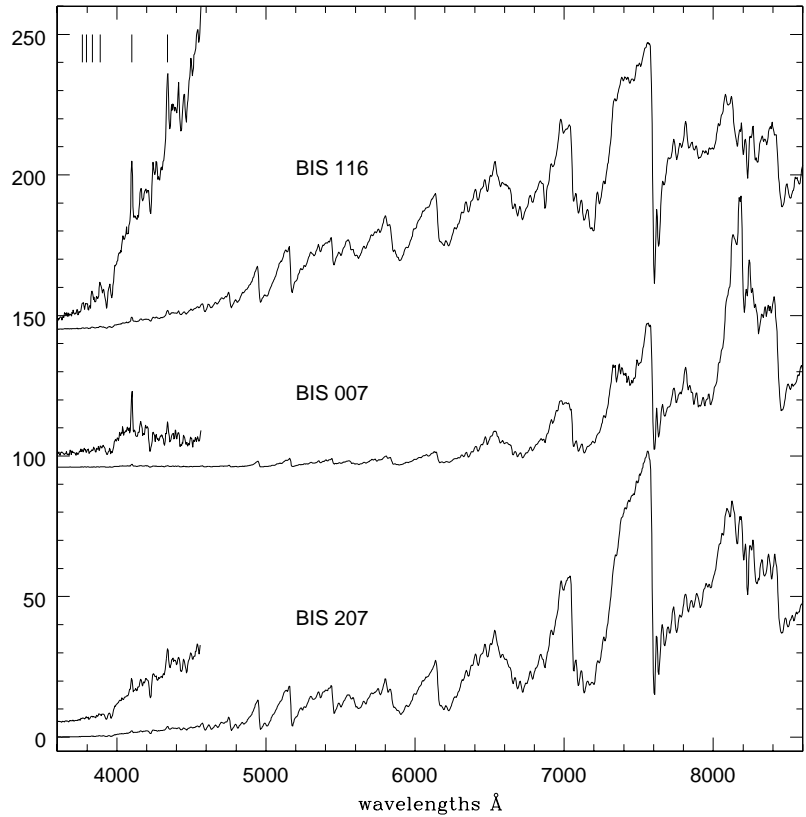}
\caption{Selected optical spectra of stars with emission lines. From bottom to top: 
BIS\,207 (M5, Semi$-$Regular);  BIS\,007 (Stype, Mira$-$type); BIS\,116 (M2$-$M4, Short$-$Period Mira$-$type). The blue part of the spectrum is replotted enlarged for better visibility above each star.  } 
\label{Fig1}
\end{figure*}
%%%

%
%   FIGURA   Selected optical spectra
%
\begin{figure*}
  \includegraphics[width=\columnwidth] {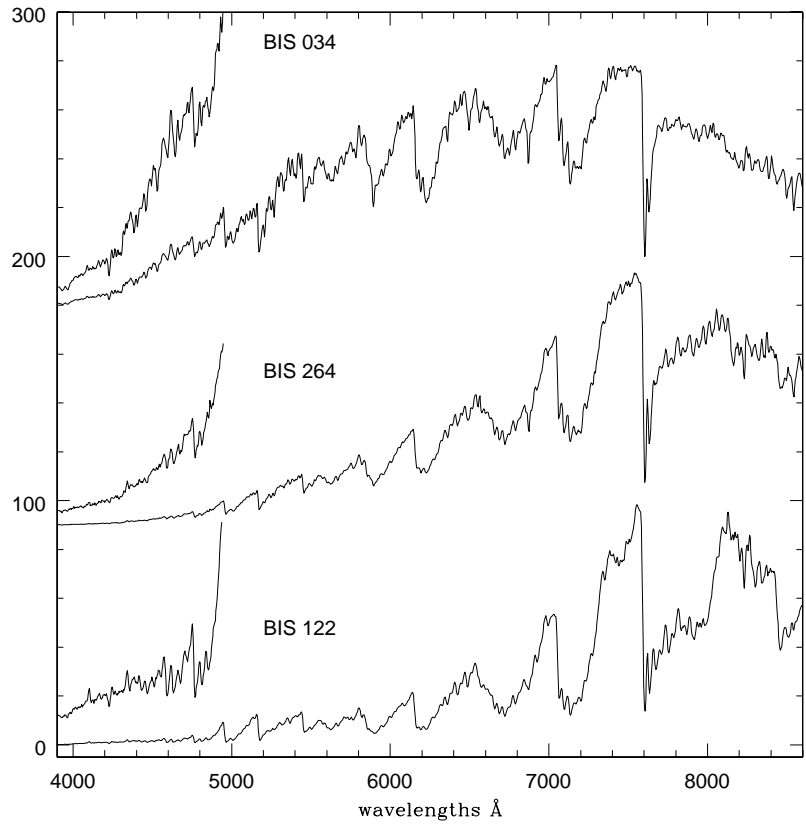}
\caption{Selected optical spectra. From bottom to top: BIS\,034 (M2, very stable); BIS\,264 (M4$+$dust); BIS\,122 (M7, Semi$-$Regular with emission lines).   The blue part of the spectrum is replotted enlarged for better visibility above each star. }
\label{Fig2}
\end{figure*}
%%%

%
%   FIGURA  spettri Carbon stars
%
\begin{figure*}
  \includegraphics[width=\columnwidth] {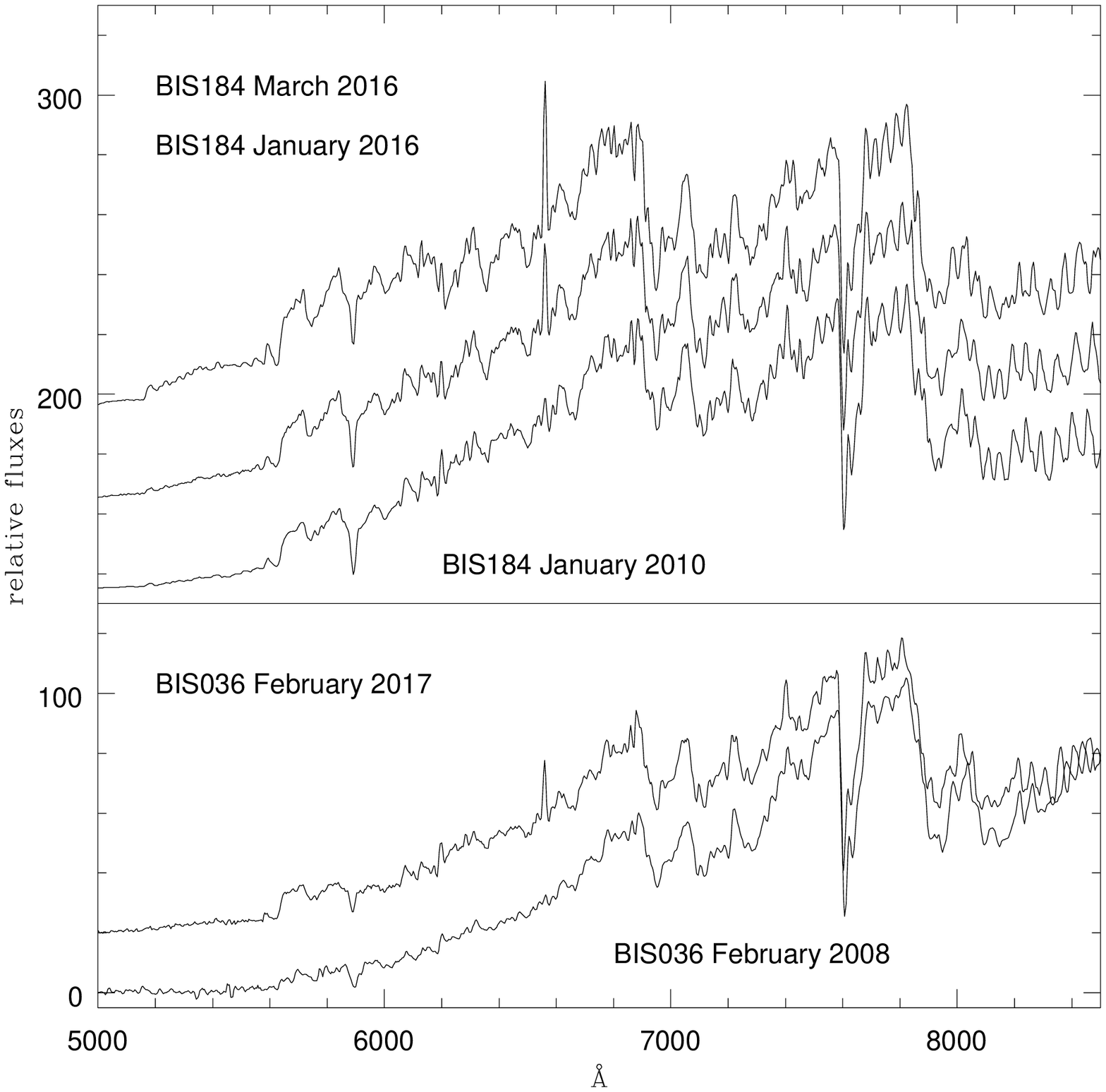}
\caption{ Spectral evolution  of the two carbon stars  BIS\,036 and BIS\,184 normalised to 100. The spectra at different dates are vertically shifted for better visibility. } 
\label{Fig3}
\end{figure*}
%%%

\medskip

{\bf Variability class~(1)} \\
{\it BIS\,007;  BIS\,116; BIS\,133 (IY~Dra); BIS\,196;  BIS\,267 }:  

\smallskip

% BIS\,007 
 BIS\,007: For this star \citet{woz04b} report an estimated period of 341 days. The light curve is compatible with that of a Mira$-$  variable star, but the photometric behaviour deserves a long term observational program to improve the accuracy of the period. The star is erroneously classified  in BIS catalog  as an N$-$type carbon star. No other spectroscopic information was found in literature. 
We observed BIS\,007 in July 2007 and July 2008. The energy distribution is similar in the two epochs as well as the  spectral features; in both epochs H$\delta$ and H$\gamma$ are in emission, other hydrogen lines being hidden by the strong molecular absorptions. We classify this object as an S-type star, similar to HD56567 (S5/6 subtype). In Fig.~\ref{Fig1} the 2008 spectrum is presented.
%The IR colors are  those of typical Long Period Variable stars (LPVs). 
% star due to the well expressed absorption bands of TiO molecule at $\lambda$ 6159, 6651, and 7100 \AA.

% BIS\,116
BIS\,116: ~~ During the ROTSE monitoring the star showed a continuous modulation with period of 160 days, in agreement with \citet {nich06}; our photometric data  are  in agreement with the expected values.  \citet{woz04b} classified this star as a Semi$-$Regular variable but the light curve
pushes toward a classification as a Short$-$Period Mira.
%In the near IR this star116 is the bluest in $J-H$ color; instead, in the Mid IR the object lies in the region of the late M stars 4.6$-$22 and  12$-$22 $WISE$ color.
We have obtained  three spectra, in 2007, 2008 and 2015. The  continuum and  the  intensity of molecular absorptions are variable. In the three spectra  the hydrogen Balmer lines are in emission, though with different intensities. In Fig.\ref{Fig1} the 2008 spectrum is presented.

% BIS\,133
BIS\,133 ( IY Dra ): ~~ This is a Mira$-$type variable star whose photographic magnitudes at minimum and maximum luminosities are presented in \citet{kaza00}. A lower limit of 351 days for the period is given by   \citet{woz04b}.
% (Fig.~\ref{lc1}).
We carefully verified our $R$ magnitudes which are all at the faint limits of the ROTSE magnitudes even taking into account the color correction.
The star is strongly saturated in the POSS red plates. The Sloan Digital Sky Survey (SDSS~\footnote{http://www.sdss.org/}) reports $r = 14.83 $~mag, corresponding to an $R$ magnitude in the range 13.80$-$14.30, depending on the adopted transformation equation; we remember anyhow that all the transformations are based on main sequence stars.
From the spectroscopic point of view IY~Dra is a very late type star very similar to the 2008 spectrum of BIS\,007.
%with strong VO molecule absorption bands in the range 8500$-$8650~\AA.
% Only the 7159 and 7861\AA ~absorption bands of TiO molecule are strong.  The atomic lines of Ti are very strong.
%The energy distribution is  the same in the two epochs.
%(see Fig. \ref{spMemi}).

% BIS\,267  
BIS\,267: ~~ The ROTSE light curve shows strong variations with continuous modulation between $R_{r}$ 10.2 and 13.2 magnitudes.
%   compatible with that of aMira$-$type star.
The bibliography based on the POSS plates gives fainter R magnitudes, while our $R$ magnitude ( 9.75 ) corresponds to a brighter object. This implies  that the period should be longer than  the 255 days reported by the automatic calculation of NSVS.
% The Sloan Digital Sky Survey (SDSS~\footnote{http://www.sdss.org/})
%did not point the field covering the coordinates of the star
%We could not find other relevant spectroscopic and  photometric information concerning  this star.\\
From the spectroscopic point of view BIS\,267 is an interesting object with the higher lines of the hydrogen Balmer series in emission ( extremely strong H$\gamma$ \, and H$\delta$, no H$\alpha$, no H$\beta$ ). The absorption spectrum is remarkable for the strength of  the TiO molecular bands

% BIS\,196
BIS\,196: ~~ The ROTSE light curve, one year long, shows strong variations with continuous
modulation between $R_{r}$ 11.0 and 12.6 magnitudes,
% compatible with that of a Mira$-$type star. 
The automatic calculation of NSVS reports a period of 313 days.
% (see Fig.~\ref{lc1}). 
\citet{step86} from objective prism spectroscopy already classified this object as an M8 type star. Our data confirm his classification. There were no emission lines in the optical spectrum  at the epoch of our observation while the shape is very similar to that of BIS\,007.
\citet{sharp95}, using the data from \citet{step86} and IR photometry, include this star in a group of possible Miras with period longer than 350 days. At the epoch of our observations the star was definitely fainter than the ROTSE minimum, even taking into account the $IRAS$ color correction.
The photometric behavior of this giant deserves a long term observational program to compute the exact period.

\medskip

{\bf Variability class~(2e)} \\
 % BIS-002, BIS\,122, BIS\,207, BIS\,264  emission line  semiregular stars
 {\it  BIS\,002; BIS\,122; BIS\,207;   BIS\,219;  BIS\,264}:  

Emission lines of the hydrogen Balmer series characterise the spectra of these stars (see Figures~\ref{Fig1} and~\ref{Fig2}); the NSVS archive classify BIS\,122, BIS\,207 and BIS\,264 as Semi$-$Regular variable with a period of 147, 181 and 154 days, respectively, but modulations with shorter periodicity and small magnitude oscillations are also present in the light curves.
Our observations show variability inside the ROTSE range. Spectral features are typical for middle to late$-$type giants.
% BIS\,002 16dec2013 R=10.91pm0.10, B=13.35pm0.07, I=8.70pm0.09

\medskip

{\bf Variability class~(2) Semi Regular} \\
% STELLE RAGGRUPPATE large amplitude semiregular, calculated quasi-periods
{\it  BIS\,043 (KP~Cam);   BIS\,198;  BIS\,276 ;
 BIS\,001; BIS\,006;  BIS\,014;  BIS\,032; BIS\,037;  BIS\,038;  
 BIS\,088 ;  BIS\,103 ;  BIS\,104; BIS\,106;
 BIS\,120 ; BIS\,123; BIS\,132; BIS\,138;  BIS\,168 ;  BIS\,173;
 BIS\,200;  BIS\,211; BIS\,213; BIS\,214; BIS\,271}:

These stars have been classified by  \citet{woz04b} as large amplitude,  Semi-Regular variables. 
 The light curves show  a semi-regular pattern with continuous modulation and large amplitude ($>$0.4 mag) variability. For  the first three stars quasi  periodicities have also been evaluated.
The spectra are  typical of M5$-$M7 giant stars.
% ROTSE observations span more than one year. 

BIS\,043 is also present as a Semi$-$Regular variable in 76th list of variable stars \citep{kaza01}. 
The visual inspection of its light curve does not support convincing evidence for the quasi periodicity of about 150 days automatically calculated by NSVS.

BIS\,138 is the only other star (besides the carbon  BIS\,184 quoted below), to have an $IRAS$ ~LRS infrared spectrum classified as 24 (star with {\it "not too thick oxigen-rich envelope"}).

BIS\,276 shows two minima at the same magnitude at the beginning and the end of the ROTSE monitoring, with a period automatically calculated of 436 days. At variance with expectation, in all our 6 observations we found the source always nearly at the same luminosity, inside the ROTSE range. 

\medskip

{\bf Variability class~(2) Irregular}. \\
% moderate amplitude irregular
 {\it BIS\,003;  BIS\,004;      BIS\,015;  BIS\,039; 
      BIS\,126;  BIS\,136;      BIS\,142;  BIS\,145; 
      BIS\,154;  BIS\,156;      BIS\,167;  BIS\,170; 
      BIS\,209;  BIS\,212;      BIS\,216;  BIS\,226:}

During the ROTSE monitoring  the light curves showed irregular variability  with maximum amplitude of 0.5 mag. Our measures are generally slightly fainter than the ROTSE values, in agreement with the expected difference due to the color correction for the spectral type of the stars, and in better agreement with the $R$ values from literature. All these stars have spectral types between M5 and M8.\\
 The light curve of BIS\,226 shows irregular variability with two minima at $R_{r}$ about 11.7 while 
our $R$ magnitude 12.85 is much fainter  in agreement with other VIZIER catalogs. 
\medskip

% small amplitude 
{\bf Variability class~(2/3)} \\
{\it  BIS\,010;  BIS\,107; BIS\,113;  BIS\,172;
      BIS\,174;  BIS\,201; BIS\,224;  BIS\,255; 
      BIS\,275;  BIS\,285:}
      
During the ROTSE monitoring these stars showed random variability with maximum excursion of 0.3 magnitudes. The $R$ magnitudes from our observations are also in agreement with the ROTSE range. Most of  these stars have an early M spectrum.

We included the M5 star  BIS\,010 in this variability class in spite of the fact that this star was neither monitored by ROTSE because of its faintness nor by the Catalina survey which does not cover its declination zone. The brightness of this star is probably overestimated in every optical catalog being the northern component of an apparent close binary (3 arcsec ). The southern component almost disappears in the POSS2 IR plate and we found that its spectrum is early G. From the comparison of the images of  DSS1 with DSS2 no magnitude variations of BIS \,010 are evident. Our  observations  in  2013, 2016, 2017 did not reveal spectroscopic or photometric variation. Its extremely red infrared magnitudes (see our companion paper)  cannot be justified by the immediate explanation of a long period variable. A  very long monitoring should be anyhow recommendable. 
 
\medskip

{\bf Variability class~(3)}. \\
% very stable stars
{\it BIS\,034;  BIS\,044;  BIS\,067;  BIS\,087;  BIS\,099;  BIS\,102;  BIS\,110;  BIS\,137; BIS\,143;  BIS\,155; BIS\,197;
 BIS\,199;  BIS\,203; BIS\,210;  BIS\,215 BIS\,228;  BIS\,247*; BIS\,248;  BIS\,256;  BIS\,258; BIS\,260 }: 

All these stars are very stable during the ROTSE monitoring  with a maximum variability of 0.1 magnitude. The photometric data available in literature and our values indicate small fluctuations compatible with ROTSE or Catalina survey data. 
No spectroscopic information was found in the literature except those by \citet{mic06}. Our revised spectral classification turned out to be early M$-$type for all stars. 
 Fig.~\ref{Fig2} shows the spectrum of BIS\,034   which  can be better classified as an S type star of subtype S2/3, very similar to  the prototypes HD 49368 and HD22649.
May be interesting is the very stable BIS\,247  showing  emission lines in the spectrum indicating a circumstellar envelope, whose origin would deserve further studies.

\medskip
 
{\bf Carbon stars } \\
{ \it BIS\,036;   BIS\,184 (HP~Cam); BIS\,222; BIS\,194}:\\
 BIS\,194 is stable while the other stars are  Irregular variables. BIS\,036 and BIS\,222 have been identified as  R Coronae Borealis candidates by different authors. These stars could not be observed by ROTSE being too faint. We made repeated observations of these stars; the first spectra obtained were presented in \citet{rossi16}. Here we will only report the spectral evolution of BIS\,036 and BIS\,184   shown in Fig. \ref{Fig3}. BIS\,194  did not change and BIS\,222 progressively faded making impossible  the acquisition of new spectra with our instruments.

For BIS\,036 we get  the light curve from the Catalina Survey
where the V magnitude ranges from 13.5 to 16.5. 
Actually we verified that on the POSS1 plate taken on February~5, 1954 the star was quite luminous, 
brighter than the nearby star having RA(2000) 05:28:56.5 and DEC(2000) +69:20:31,
while on the POSS2 plate taken on October~31, 1994 it appears much fainter,  likely after an episode of mass ejection.
On the FBS plate 0138a taken on November~21, 1969 the star appears as a very short faint segment, consistent with R$\le$16. 
  On the basis of the Catalina light curve and of  the energy distribution,  recently \citet{lee15} included this star  in a list of R~Coronae Borealis  candidates, although the strong photometric variations, typical for R CrB stars have never been reported. 
We have observed this star in 2008, 2016 and 2017   detecting  changes in the  photometry and spectrum.
The spectrum of BIS\,036 is typical for a very late carbon star : in addition to the CN bands, the (0,1) transition at $\lambda$ 5636 \AA \ (Swan System) of the \ion{C$_2$}\,molecule is barely visible.  A deep NaD absorption is present at $\lambda$ 5895 \AA \ possibly produced in the circumstellar envelope. The 2016 and 2017 spectra are almost identical, so we present only the last one to avoid  confusion. An important difference  between the 2008 and 2016 spectra is the appearance in 2016 of a strong  H$\alpha$ in emission: this feature, added to the photometric behavior, while compatible with a long period variable star rules out the R CrB hypothesis.

%BIS 184
BIS\,184:  This N type carbon star  is a known Semi$-$Regular variable star (HP Cam),  listed in the General Catalog of Variable Stars \citep{Sam17}.  A photometric variability with a period of 296 days is reported by \citep{woz04b}.
% (GCVS~\footnote{http://www.sai.msu.su/gcvs/gcvs/}):
% (GCVS version 2010)
 The Low Resolution Spectrum from $IRAS$ ~is the only spectroscopic  reference \citep{kwok97} where this object is classified as a carbon star based on the presence of the SiC emission feature at 11.2~$\mu$m (LRS classification is 44, according to \citealp{little87}).
This classification is confirmed by our optical spectra which also showed variations, as expected  from its classification.  A substantial  strengthening  of H$\alpha$ occurred between January and March 2016. In  Fig.~\ref{Fig3} we do not show the February 2017 spectrum, which is practically the same as January 2016.

The variability of  BIS\,222, was  ascertained from the comparison of the historical plates of  the POSS and the FBS: it was as strong as the two nearby stars in the Red POSS1 plate
taken in 1955, when its red magnitude was about 17.3. It is barely visible in the FBS plate No. 1332 (1975): taken into account that 17.5~mag is the limit in the photographic band for the FBS plates, this could be a clue for the magnitude at that epochs. The star is invisible in  the POSS2 red plate (1996) and very strong in the POSS2 photographic IR (1997) plate; on January 2010 we have measured an $R$ magnitude of 18.5. For all these reasons we assigned to  BIS\,222 the variability class~{\bf (2)}.
% This is the reddest carbon star  in all the color$-$color diagrams.
Our spectrum is consistent with that of a dust$-$enshrouded carbon star of a very late (N8$-$N9) subtype, similar to the well known IRC~$+$10216 (CW~Leo), an archetype of post AGB star and Pre$-$Planetary Nebula object. The only spectral features visible in the optical spectrum are the red and Near IR bands of \ion{CN}\,molecule at $\lambda$~6952, 7088 and 7945~\AA, 8150~\AA.
 H$\alpha$ is present in  emission ( see Fig.1 of \citealp{rossi16} ).
 %(see Fig.~\ref{spNdust}).
 Note that  \citet{tisse12} included this star among the R CrB candidates on the basis of the infrared colors and the energy distribution ( star No. 1542 ), but  the presence of H$\alpha$ could make questionable this classification. Being the star fainter during our more recent observations,  we could  obtain photometric data  only.
%  http:// www.mso.anu.edu.au/~tisseran/RCB/    

% BIS\,194
{\it BIS\,194}: ~~ the short monitoring (120 days only) by ROTSE  indicates a  very small modulation between 9.4 and 9.6~mag. Our $R$ magnitude, 9.3 $\pm$ 0.2, is in good agreement with the value indicated by ROTSE. We tentatively assign the variability class~{\bf (3)} taking into account the ROTSE monitoring and our photometry only. No spectroscopic information were found in the literature. Our spectrum is that of an N type carbon star ( see Fig.1 of \citealp{rossi16} ) with well expressed absorption of \ion{C$_2$}\, molecule (Swan system) with band heads at $\lambda \lambda$ 4737, 5165, 5636, 6122 and 6192~\AA. The \ion{CN}\,molecule is also present with the bands at $\lambda \lambda$\, 5264, 5730, 5746, 5878, 6206, 6360, 6478, 6631, 6952, 7088, 7259, 7876$-$7945 and 8150~\AA. In the three  FBS plates  containing the  star, the blue molecular bands are  well visible  but the red part of the spectrum is saturated.
%%%%%%%%%%%%%%%%%%%%%%%%%%%%%%%%%%

\section{Physical parameters}    \label{PP}

 Having defined the above mentioned data, we are now in a better position to discuss  the results from the optical photometric behavior and the spectral types of the stars. Where possible, we  derive absolute magnitudes, distances and mass loss rates by applying different methods, according to the different types and characteristics of our objects. 
These parameters are commonly derived from empirical relations based on stars of well known distances: the mass loss rate determination is the most uncertain.
In fact the various models generating dust, wind and circumstellar shells, are  based on similar hydrodynamical equations but on  different hypothesis of gas/dust ratios and different time dependence.

Several stars in our sample are variable; in these cases calculations to obtain physical parameters should imply time$-$averaged IR magnitudes. To this purpose we have used  2MASS and $IRAS$ ~catalogs. Being aware that results from sparse observations must be considered with care, but, having a limited phase coverage of the measurements, obtained at random phases, we used  the data as if they were the mean values of the magnitudes and, as uncertainties, we adopted the typical amplitude.

\smallskip

{\bf Carbon stars}\par\noindent

To estimate the absolute luminosity and distance of the ``naked'' carbon star BIS\,194, we used the relation between the color index $J-K$ and the absolute $K$ magnitude M$_K$ obtained by \citet{dem07}, valid for $J-K$ in the range between 1.4 and 2.3 magnitudes. \\
We obtained  M$_K = -7.88$~mag and d$=$3.8~kpc.\\
Concerning  the N$-$type dusty carbon stars  BIS\,036, BIS\,184 and BIS\,222, we applied various empirical relations and models  available in literature. The results are in general agreement except for the mass loss rates, where the model dependence is strong and the differences are significant from one method to another. 
Our procedures are described below and the results are summarized in Table~\ref{tab:tabcarb} where the columns have the following meaning:
1~- BIS number;
2~- absolute $K$ magnitude M$_K$;
3~- range of distances d;
4~~ derived distance to the galactic plane Z
5~- apparent bolometric magnitude m$_{bol}$;
6~- absolute bolometric magnitude M$_{bol}$;
7~- mass loss from \citet{lebe98};
8~- mass loss from \citet{wfm06}.  

We have derived the range of absolute $K$ magnitudes using the relations between M($K_s$) and ($J-K_s$)$_o$ from the calibration by \citet{mau08}. From these values we then computed the range of distances to the stars. 
%The results are reported in columns~2 and 3 of Table~\ref{tabcarb}.

%    \caption{A Simple Table (\lowercase{$x=1.0$})\tabnotemark{a}} \label{tab:ion_ab}
%% TABLE  carbon rich m$IRAS$
%
\setcounter{table}{1}
\begin{table*}
\caption[] {Physical parameters and distances for the dusty carbon stars.}
\label{tab:tabcarb}
\begin{flushleft}
\begin{tabular}{cccccccc}
\hline
\noalign{\smallskip}
BIS     &  M$_K$      &      d  & Z    &  m$_{bol}$  &  M$_{bol}$   &  Log $\dot{\rm M}$ & Log $\dot{\rm M}$ \\
No.  &  mag        &     kpc     &  kpc &   mag     &     mag      &        LW-98         &       Wh-06         \\
\hline     
        &             &             &             &       &       &                    &               \\
036 & -6.9\,:\,-7.3 & 10.7\,:\,13.0 &   3.4\,:\,4.0    &10.6\,:\,11.0 & -4.2\,:\,-4.9  &   -5.6   &  -5.1  \\
184 & -7.5\,:\,-7.8 &  1.5\,:\,1.8   &    0.6\,:\,0.7    & 6.8\,:\,6.9   & -4.1\,:\,-4.3  &  -6.3  &  -5.8  \\
222  & -7.0\,:\,-7.4 &  9.8\,:\,12.6 &    3.0\,:\,3.8    &9.9\,:\,11.0 & -4.4\,:\,-5.5  &  -5.3  &  -4.9    \\
\noalign{\smallskip}
\hline
\noalign {\smallskip}
\end{tabular}
\end{flushleft}
\end{table*}
%

%  
%% TABLE  Oxygen Rich M$IRAS$
%
%\setcounter{table}{3}
\begin{table*}
\caption[] {Physical parameters and distances of the O$-$rich Miras.}
\label{tab:tabmira}
\begin{flushleft}
\begin{tabular}{cccccccc}
\hline
\noalign{\smallskip}
BIS       &   P    & M$_K$ & m$_{bol}$ & M$_{bol}$ &   d & Z     &  Log $\dot{\rm M}$  \\
No.    &  days  &  mag  &     mag   &     mag   & kpc & kpc    &                     \\
\hline
          &        &       &           &           &         &      &                     \\
007 &  341   & -7.87 &  7.3   & -4.80   & 2.65 $\pm$.40 &   0.6 : 0.8   & -7.92 $\pm$0.35  \\
116  &  161   & -6.65 & 9.7   &   -3.81   & 5.00 $\pm$.55  &  3.2 : 3.9  & -7.43 $\pm$0.28 \\
133  & $>$351 & -7.90 &  9.1   &  -4.84   & 5.90 $\pm$.85   &  2.3 : 3.0  & -7.26 $\pm$0.40     \\
196  &  316   & -7.72 &     7.9   &   -4.69   & 3.15 $\pm$.45  &  1.6 : 1.8   & -7.35 $\pm$0.45     \\
267  &  $>$255   & -7.40 &   9.1 & -4.45   & 5.20 $\pm$.85   &  2.5 : 3.5  & -8.2 $\pm$0.5     \\
\noalign{\smallskip}
\hline
\noalign {\smallskip}
\end{tabular}
\end{flushleft}
\end{table*}

A cross check of the estimated absolute $K$ magnitude can be made in the case of BIS\,184 using the cited period of 296 days.

\smallskip\noindent
From the relation  by \citet{knapp03}:\\
 M$_K= -1.34$ LogP$(d)\, -\, 4.5 \longrightarrow $ \\
~M$_K= -7.8 \pm 0.4$~mag.

\smallskip\noindent
From the  calibration by \citet{white12}: \\
 M$_K= -7.18-3.69\,($LogP$\, -2.38)  \longrightarrow $ \\
 ~M$_K = -7.5 \pm 0.3$~mag. 
  
 \smallskip\noindent
Both values are in good agreement with those obtained using the \citet{mau08} calibration reported in Table~\ref {tab:tabmira}.

We have  computed the apparent bolometric magnitude\\
    $m_{bol} = K$ + BC$_K$ 
 ~~using  the calibrations to the bolometric corrections  by  \citet{wfm06}.
%  BC$_K$ can be calculated as   a function of several IR colors   by applying the  polynomial fittings provided by \citet{wfm06}. 
We  obtained a range of values inside the limits reported in column~4 of Table~\ref{tab:tabcarb}.
The results obtained using the calibrations by \citet{lebe01} are in good agreement with those listed in the table.

From the  derived distances and from m$_{bol}$  ~ we could infer the range of absolute bolometric magnitudes. 

For BIS\,184 M$_{bol}$ can also be obtained in an independent way, using the period$-$luminosity relation by \citet{feast06} (see their Fig.~2 and eq.~1 with c$=$2.06). The result, M$_{bol} =$ $-$4.22 $\pm$ 0.24 mag. is in perfect agreement with that  reported in Table~\ref{tab:tabcarb}.
 
We have finally estimated a range of mass loss rates being aware that the uncertainties are quite large. The combinations between several IR colors and mass loss rate was studied by \citet{lebe98} starting from stationary and from time$-$dependent models, with fixed assumptions on the parameters of the circumstellar shells; for the stars in common with a paper by \citet{white94} they obtained a satisfactory agreement, though with lower values of $\dot{\rm M}$.
A good correlation between $K-[12]$ color and the total mass loss rate was derived by \citet{wfm06} refining the previous work.  Our results confirm the systematic difference of a factor of three between the two methods. In the last two columns of Table~\ref{tab:tabcarb} we report the results (in M$\odot$/year), the uncertainties are about $\pm$~0.2 in the logarithm.

\smallskip

{\bf O$-$rich Miras  }\par\noindent
%%  MIRA type stars magnitudes distances and mass loss
For the M$-$type Mira variables a number of relations involving period and IR luminosities (discussed in our companion paper)  can be applied. The results are summarised  in Table~\ref{tab:tabmira} where the meaning of the columns is the following: 
1~- BIS number; 
2~- variability period P reported by NSVS; 
3~- absolute $K$ magnitude M$_K$; 
4~- apparent bolometric magnitude m$_{bol}$; 
5~- absolute bolometric magnitude M$_{bol}$; 
6~- distances and corresponding errors; 
 7~~ derived distance to the galactic plane; 
8~- mass loss.

The results for BIS\,133 reported in the table are lower limits, being the period longer than 351 days.
Similarly we have  assigned class $1$ to BIS\,267 although this star is not known to be a Mira type but shows a regular pattern in the light curve of ROTSE experiment, as described above. 
At the opposite side, BIS\,116 has the shortest period, the smallest amplitude and the earliest spectral type: its position in almost all color$-$color diagrams is aligned with the long-period Miras.
%, as one can see in Fig.~\ref{lc1},$IRAS$
\citet{wmf00} divided the Mira with period below 225 days in two groups, ``Short Period$-$Blue'' and ``Short Period$-$Red''  depending on their infrared colors and average spectral types. All spectroscopic and photometric characteristics of BIS\,116 lead us to place this star in the  group of the ``Short Period$-$Blue'' Miras.

We have computed the absolute $K$ and bolometric magnitudes using the relation between period and magnitude given in \citet{white12} and references therein: 

 \smallskip
 
 M$_K$ = -3.69* (Log(P) $-$ 2.38) $-$ 7.33 
  
 M$_{bol}$= -3.00* Log(P) $+$ 2.8
\smallskip

For the uncertainty on the period, after accurate inspection of the NSVS light curves, we assumed five days for BIS\,116 and ten days for the other stars. By propagating the errors we obtained $\Delta$M$\sim$0.1~mag for BIS\,116 and  BIS\,267 and $\Delta$M$\sim$0.07~mag for the other stars. The uncertainties are practically the same for M$_{bol}$ and M$_K$.

We  have also computed the apparent bolometric magnitude with BC$_K$ $=$ 3.15~mag  for BIS\,007, BIS\,133, BIS\,196,  BC$_K$ $ =$ 2.9~mag   for BIS\,267, and BC$_K$ $ =$ 2.8~mag   for BIS\,116, derived from the calibrations by \citet{wmf00}.
As uncertainties we assumed the typical amplitude of the $K$ magnitude that 
is 0.4~mag for the late$-$type Miras and 0.2~mag for BIS\,116  and BIS\,267.
    
We could then infer a crude estimate of the distances and of the mass loss rate.
The distances derived from the bolometric and the $K$ magnitudes agree very well within the errors.
To compute the mass loss we used calibration between $\dot{\rm M}$ and $K-$[12] 
color index obtained by \citet{lebe98}. Here the contribution to the uncertainties is also due to the 
non$-$simultaneous observations of the different sets of IR data.

\smallskip

{\bf Semi$-$Regular variables  }\par\noindent
%%  semi regular stars magnitudes distances and mass loss 
Our sample includes several Semi$-$Regular variables, but only five, namely  BIS\,122, BIS\,198, BIS\,207, BIS\,209 and BIS\,276 have well sampled light curve in the NSVS.
We could derive a range of absolute magnitudes and distances by applying the relations found by \citet{knapp03} and by \citet{barthes99}. These last authors found different relations P$-$M$_K$ for different kinematics characteristics and obtained significant results by dividing their data into four groups, representative of four different populations. 
Group~1 and 2  have kinematics characteristics corresponding to old disk stars; group~3 has kinematics indicating a younger population, group 4 contains only high velocity stars.  Only group 1 contains long periods. 

We report the results  in Table~\ref{tab:tabSR} where the columns have the following meaning:
1~- BIS number;
2~- variability period P;  
3~- $K$ magnitude from 2MASS; 
4~- absolute $K$ magnitude M$_K$ from \citet{knapp03};  
5~- range of M$_K$ from \citet{barthes99}, obtained applying the relation giving the best agreement with \citet{knapp03};  
6~- corresponding group number following  \citet{barthes99}; 
7~- range of distances from the minimum and maximum of columns~4 and~5. 
 8~~ derived distance to the galactic plane.

 %    
%% TABLE 5
%
% \setcounter{table}{1}
\begin{table}
\caption[] {M(K) and distances for 5 Semi$-$Regular variable stars.}
\label{tab:tabSR}
\begin{flushleft}
\begin{tabular}{cccccccc}
\hline
\noalign{\smallskip}
BIS   &  P   & $K_s$ &  M$_K$  &    M$_K$    & gr  &   Dist$_K$ & Z  \\
      & days &  mag  &   K03   &     B99     & B99 &     kpc  & kpc   \\
\hline
      &      &       &         &             &     &     &        \\
122   & 147  & 6.63  &  -7.40  & -7.35:-7.44 &  2/3  &  6.3 : 6.5&  3.5 : 3.6 \\
198   & 161  & 5.71  &  -7.45  & -7.42:-7.53 &  2/3  &  4.2 : 4.4& 2.5 : 2.6 \\
207   & 181  & 6.42  &  -7.52  & -7.52:-7.65 &  2/3 &  6.1 : 6.5 & 3.9 :  4.1     \\
209   & 296  & 5.00  &  -7.80  & -7.95:-8.12 &  2/3 &  3.6 : 4.2 &  2.3:  2.6    \\
276   & 436  & 6.16  &  -8.00  & -7.94:-8.18 &    1  &  6.5 : 7.4 &  3.4 : 3.9    \\
\noalign{\smallskip}
\hline
\noalign {\smallskip}
\end{tabular}
\end{flushleft}
\end{table}

\smallskip

{\bf Non variable stars  }\par\noindent
%%  stable stars magnitudes distances and mass loss 
All the non$-$variable or small amplitude variable stars are  giant of  M0-M4 sub-classes.
For these stars we derived a range of distances between 1.0 and 3.3~kpc, by adopting  absolute visual magnitudes  $-$0.7  $\le$ M$_V$ $\le -$1.6  (\citet{joque86}; \citet{the90}; \citet{sparga00})
and averaging the apparent visual magnitudes retrieved from the catalogs UCAC4 \citep{zac12} and GSC2.3. Considering  the galactic latitude yields to distances to the galactic plane  ranging from thick disk to halo, ( 0.45$-$2.1 kpc, $ \overline{Z}=1.03 , \sigma=0.45$\,kpc).  
 %( 0.5$-$2.1 kpc) 
Regarding the supergiant BIS\,137 (M0~I), by adopting an absolute visual magnitude  M$_V  \sim $\,5.0,  we estimate a distance $\sim$\,6 kpc, and Z$\sim$\,2.6 kpc. Many stars are likely within the reach of a direct parallax measure by GAIA.

%  \begin{figure*}
%   \includegraphics[width=\columnwidth]{Fig4.eps}
 %  \caption{  Histogram  of the distances of the observed stars to the galactic plane.   } 
%   \label{Fig4}
%   \end{figure*}

\section{Concluding remarks}   \label{Conc}

In this work we have studied spectroscopic and photometric optical characteristics of a sample 
 (95) of  late type stars from the BIS catalog. The original situation  of the entire catalog (276 stars) included 30\% unclassified objects and 55\% with very uncertain classification.

From our new spectroscopic data we have revised the spectral classification of the observed targets: in 45\% of the cases we have deeply modified or improved the previous classification.
%which was based on low resolution spectroscopy or on photometric criteria only.
 Four objects came out to be carbon N$-$type stars, nine are earlier than M,
  the others are M$-$type giants.

We have divided our stars into three main variability classes: regular variable,  irregular variable, photometrically stable. About 60\% of the stars show large amplitude, irregular or semi$-$regular light curves; in this category we have included the three dust enshrouded carbon stars. 
All the early M$-$type stars and the naked carbon star BIS\,194 were found to be stable. Only 5 stars are Mira variables.
 Our study pointed out some peculiar stars which deserve more detailed studies.

 Our spectral classification together with the data  collected from literature allowed us to  estimate absolute magnitudes, mass loss and distances for  a number of targets, finding a good agreement between results obtained from different methods.
 
Knowing the distances to the Sun and the galactic latitude, we can infer that the median distance of the variable stars  from the galactic plane is 3.0~kpc, ($\sigma$=1.5) none being farther than 5.4~kpc.
 Our spectral resolution does not allow to investigate  for chemical differences between thick disk  and halo stars.  Anyhow, given the distance from the galactic plane, most of these stars are likely to be halo members.
 
 For the non$-$variable stars we derived an average distance of 1.03 kpc from the galactic plane, suggestive of a mixed population of thick$-$disk and halo.
 
Regarding the evolutionary status of our sample, most of the stars have an absolute magnitude appropriate for being in the AGB phase. 

 In a companion study  (Gaudenzi et al.,  submitted) we   have analysed   the Infrared properties of all the stars of the BIS catalog. To this purpose  
we have made use of  near-IR (2MASS), mid-IR ($WISE$), and far- IR ($IRAS$ ~and $AKARI$) photometric data to investigate their behaviour on various color-magnitude and color-color diagrams and graphically distinguish various types of sources.

\begin {acknowledgements}
{\bf Acknowledgements.}
This research has made use of the SIMBAD database, operated at CDS, Strasbourg, 
France.
This publication has made use of data products from:
 the Two Micron All$-$Sky Survey database, which is a joint project of the University of Massachusetts and the Infrared Processing and Analysis Center/California Institute of Technology;
the Wide$-$field Infrared Survey Explorer, which is a joint project of the University of California, Los Angeles, and the Jet Propulsion Laboratory/California 
Institute of Technology, funded by the National Aeronautics and Space 
Administration; 
the Northern Sky Variability Survey (NSVS) created jointly by the Los Alamos National Laboratory and University of Michigan;
 the NASA/IPAC Extragalactic Database (NED) which is operated by the Jet Propulsion Laboratory (JPL), California Institute of Technology, under contract with the National Aeronautics and Space 
Administration;
the International Variable Star Index (VSX) database, operated at AAVSO, Cambridge, Massachusetts, USA.
The University ``La Sapienza'' of Rome, Italy, supported the project with funds from MIUR.
\end{acknowledgements}

\setcounter{table}{0}
\begin{table*}\footnotesize\centering
 \setlength{\tabnotewidth}{0.5\columnwidth}
 \tablecols{7}
\caption{Log of  observations, spectral classification, magnitudes and variability  class.  \tabnotemark{a} }
\label{tab:tab1}
%\begin{tabular}{@{}lccllcl@{}}
\setlength{\tabnotewidth}{0.95\linewidth}
\begin{tabular}{lccllcl}
\toprule
BIS  & $IRAS$ FSC    &    Date   & Sp. type   &  $R$ mag~             & $R_{r}$     & Var. class \\
\hline
     &             &           &                       &             &            \\
001  & F03503+6918 & 29 Nov 13 & M7  III       & 10.17 $\pm$ 0.95~ $^1$     &             & 2     \\
     &             &           &            &                       &             &       \\
002  & F03535+6945 & 16 Dec 13 & M5 III+ em    & 10.9 $\pm$ 0.10~ $^1$     &             & 2e    \\
  &  & 13 Jan 16 &    & 10.92 $\pm$ 0.08~ $^1$     &             &    \\
  &  & 13 Jan 16 &    & 8.7 $\pm$ 0.1~ $^1$     &             &   I mag  \\
     &             &           &            &                       &             &       \\
003  & F03558+7007 & 16 Dec 13 & M6 III         &  9.78 $\pm$ 0.08~ $^1$      &             & 2     \\
     &             &           &            &                       &             &       \\
004  & F03564+7148 & 16 Dec 13 & M3  III         &  9.93 $\pm$ 0.06~ $^1$      &             & 2     \\
     &             &           &            &                       &             &       \\
006  & F04067+7139 & 17 Dec 13 & M6 III          & 10.26   $\pm$ 0.10~ $^1$ &            & 2     \\
     &             &           &            &                       &             &       \\
007  & F04125+7106 & 22 Jan 08 & S5/6 & 11.25 $\pm$ 0.06~$^1$
    &  9.3$-$11.8 & 1     \\
     &             &           &            &                       &             &       \\
010  & F04137+7016 & 18 Dec 13 & M5III-IV   & 16.77$\pm$ 0.05~ $^1$ &             & 2/3    \\
  &  & 14 Feb 17 &    & 16.63$\pm$ 0.05~ $^1$ &             &     \\
    &  & 14 Feb 17 &    & 14.01$\pm$ 0.07~ $^1$ &             & I mag.     \\
     &             &           &            &                       &             &       \\
014  & F04173+7232 & 18 Dec 13 & M6 III          & 10.93    $\pm$ 0.07~ $^1$             &             & 2     \\
     &             &           &            &                       &             &       \\
015  & F04174+7111 & 18 Dec 13 & M7 III          & 11.43  $\pm$ 0.12~ $^1$               &             & 2     \\
%     &             & 18 Dec 13 &            & 11.31  $\pm$ 0.07~ $^1$               &             &       \\
     &             &           &            &                       &             &       \\
032  & F05088+6948 & 17 Feb 14 & M8  III         & 12.51$\pm$ 0.05~ $^1$ &             & 2   \\
     &             &           &            &                       &             &       \\
%034 & F05143+7048 & 11 Feb 07 & M2 III     &                       &             & ~     \\
034  & F05143+7048 & 22 Jan 08 & S2/3   &  8.3  $\pm$ 0.1 ~$^1$ &  8.6$-$8.8  & 3     \\
     &             &           &            &                       &             &       \\
%036 & F05235+6918 & 11 Feb 07 & N+dust     &                       &  ~          &       \\
036  & F05235+6918 & 22 Jan 08 & N+dust     & 16.45 $\pm$ 0.05~$^1$ &  ~          & 2     \\
  &  & 13 Jan 16 &      & 15.53 $\pm$ 0.08~$^1$ &  ~          &     \\
  &  & 13 Jan 16 &      & 18.27 $\pm$ 0.09~$^1$ &  ~          & V mag     \\
   &  & 18 Mar 16 &      & 16.1 $\pm$ 0.1~$^1$ &  ~          &      \\
    &  & 14 Feb 17 &      & 15.30 $\pm$ 0.08~$^1$ &  ~          &      \\
     &             &           &            &                       &             &       \\
037  & F05286+7105 & 18 Dec 13 & M7  III   & 12.72   $\pm$ 0.08~ $^1$    &     & 2\\
     &             &           &            &                       &             &       \\
038  & F05342+7120 & 12 Feb 07 & M8 III     & 11.48 $\pm$ 0.03      & 11.1$-$11.6 & 2     \\
     &             &           &            &                       &             &       \\
%039 & F05357+7054 & 11 Feb 07 & M3 II      &                       &  ~          &       \\
039  & F05357+7054 & 22 Jan 08 & M3 II      &  8.9  $\pm$ 0.2~ $^1$ &  9.2$-$9.4  & 2    \\
     &             &           &            &                       &             &       \\
%043 & F05435+6908 & 11 Feb 07 & M6 III     &                       &  ~          &       \\
043  & F05435+6908 & 22 Jan 08 & M6III      & 10.93 $\pm$ 0.03~$^1$ & 10.2$-$10.8 & 2     \\
     &             & 07 Dec 08 &            & 10.47 $\pm$ 0.04~$^2$ &  ~          &       \\
     &             &           &            &                       &             &       \\
\bottomrule
 \tabnotetext{a}{
~$^1$ Loiano Observatory; $^2$ Cima Ekar Observatory;  all the other data were obtained with the TACOR telescope. }
\end{tabular}
\end{table*}

%%%%%%%%%%%%%%%%%%%%%

\setcounter{table}{0}
\begin{table*}\footnotesize\centering
 \setlength{\tabnotewidth}{0.5\columnwidth}
 \tablecols{7}
\caption{Log of observations, spectral classification,  magnitudes and variability   class.  C\lowercase{ontinue} }
\label{tab:tab1}
%\begin{tabular}{@{}lccllcl@{}}
\setlength{\tabnotewidth}{0.95\linewidth}
\begin{tabular}{lccllcl}
\toprule
BIS  & $IRAS$ FSC    &    Date   &  Sp. type   &  $R$ mag~             & $R_{r}$     & Var. class \\
\hline
     &             &           &             &                       &             &       \\
044  & F05468+7300 & 17 Dec 13 & M4  III    &  9.12  $\pm$ 0.1~ $^1$               &             & 3     \\
     &             &           &            &                       &             &       \\
%067 & F05504+6215 & 11 Feb 07 & M1 III     &                       &  ~          &       \\
067  & F05504+6215 & 22 Jan 08 & M1 III  &  9.2  $\pm$ 0.2~ $^1$ &  9.2$-$9.4  & 3  \\
     &             &           &            &                       &             &       \\
087  & F06404+6324 & 18 Dec 13 & M1 III     &  9.22 $\pm$ 0.2~ $^1$ &             & 3     \\             
     &             &           &            &                       &             &       \\
088  & F06408+6402 & 18 Dec 13 & M3 III          & 10.07 $\pm$ 0.05~ $^1$  &             & 2   \\
     &             &           &            &                       &             &       \\
099  & F08209+6303 & 17 Dec 13 & M3  III         & 10.09  $\pm$ 0.08~ $^1$      &      & 3 Catalina  Var \\
     &             &           &            &                       &             &     \\
102  & F08293+6131 & 17 Dec 13 & M3.5 III        &  9.5     $\pm$ 0.1~ $^1$    &      & 3     \\
     &             &           &            &                       &             &       \\
103  & F08588+6442 & 17 Feb 14 & M5  III         & 10.73 $\pm$ 0.05~ $^1$ &             & 2     \\
     &             &           &            &                       &             &       \\
104  & F09329+6230 & 16 Dec 13 & M6 III          & 11.19 $\pm$ 0.05      &     & 2 Catalina  \\
     &             &           &            &                       &             &      \\
106  & F10009+6459 & 20 May 14 & M5 III     &  9.86    $\pm$ 0.09~ $^1$             &             & 2  Catalina \\
     &             &           &            &                       &             &     \\
107  & F10023+6334 & 21 May 14 & M3.5 III   &  9.7     $\pm$ 0.1~ $^1$            &             & 2/3     \\
     &             &           &            &                       &             &       \\
110  & F13187+6237 & 21 May 14 & M0 III     &  8.5   $\pm$ 0.2~ $^1$              &             & 3     \\
     &             &           &            &                       &             &       \\
113  & F14492+6231 & 21 May 14 & M7 III     & 10.73      $\pm$ 0.05~ $^1$           &             & 2/3  Catalina \\
     &             &           &            &                       &             &     \\
116  & F15181+6241 & 13 Jul 07 & M0--M4 III & 11.0  $\pm$ 0.1~ $^1$ & 10.2$-$11.3 & 1 Catalina     \\
     &             & 07 Jul 08 &            & 11.4  $\pm$ 0.1~ $^1$ &  ~          &       \\
     &             & 06 Jul 11 &            & 10.9  $\pm$ 0.1       &  ~          &       \\
     &             & 02 Aug 11 &            & 10.6  $\pm$ 0.1       &  ~          &       \\
     &             & 11 Aug 11 &            & 10.3  $\pm$ 0.3       &  ~          &       \\
     &             & 22 Oct 11 &            & 11.0  $\pm$ 0.1       &  ~          &       \\
     &             & 28 Mar 15 &            & 11.3  $\pm$ 0.1~ $^1$     &  ~          &       \\
     &             &           &            &                       &             &     \\
120  & F16359+6439 & 12 Jul 07 &  M6 III     & 11.20 $\pm$ 0.07~$^1$ & 10.0$-$10.6 & 2     \\
     &             & 06 Jul 11 &             & 10.87 $\pm$ 0.06      &  ~          &       \\
     &             & 05 Aug 11 &             & 11.27 $\pm$ 0.05      &  ~          &       \\
     &             & 16 Aug 11 &             & 11.20 $\pm$ 0.06      &  ~          &       \\
     &             & 09 Oct 11 &             & 10.80 $\pm$ 0.07      &  ~          &       \\
     &             &           &             &                       &             &       \\  
122  & F17249+6428 & 12 Jul 07 &  M7 III     & 12.41 $\pm$ 0.05~$^1$ & 11.4$-$12.4 & 2e Catalina  \\
     &             & 01 Aug 07 &             & 12.48 $\pm$ 0.05~$^1$ &  ~          &       \\
     &             & 10 Jul 11 &             & 12.30 $\pm$ 0.07      &  ~          &       \\
     &             & 06 Aug 11 &             & 12.48 $\pm$ 0.05      &  ~          &       \\
     &             & 22 Aug 11 &             & 12.44 $\pm$ 0.05      &  ~          &       \\
     &             & 01 Oct 11 &             & 13.06 $\pm$ 0.05      &  ~          &       \\
     &             &           &             &                       &             &       \\
\bottomrule
 \end{tabular}
\end{table*}

%%%%%%%%%%%%%%%%%%%

\setcounter{table}{0}
\begin{table*}\footnotesize\centering
 \setlength{\tabnotewidth}{0.5\columnwidth}
 \tablecols{7}
\caption{Log of  observations, spectral classification,     magnitudes and variability   class.  C\lowercase{ontinue} }
\label{tab:tab1}
%\begin{tabular}{@{}lccllcl@{}}
\setlength{\tabnotewidth}{0.95\linewidth}
\begin{tabular}{lccllcl}
\toprule
BIS  & $IRAS$ FSC    &    Date   & Sp. type   &  $R$ mag~             & $R_{r}$     & Var. class \\
\hline
     &             &           &            &                       &             &       \\
123  & F17305+6432 & 03 Aug 07 &  M5 III     & 10.8  $\pm$ 0.1~ $^1$ & 10.5$-$11.0 & 2     \\
     &             & 10 Jul 11 &             & 10.6  $\pm$ 0.1       &  ~          &       \\
     &             & 05 Aug 11 &             & 10.30 $\pm$ 0.05      &  ~          &       \\
     &             & 22 Aug 11 &             & 10.38 $\pm$ 0.07      &  ~          &       \\
     &             & 01 Oct 11 &             & 10.18 $\pm$ 0.06      &  ~          &       \\
     &             &           &             &                       &             &       \\
126  & F17579+6118 & 04 Aug 07 &  M5 III     & 10.08 $\pm$ 0.08~$^1$ &  9.7$-$10.0 & 2  Catalina    \\
     &             & 09 Jul 11 &             & 10.0  $\pm$ 0.1       &  ~          &       \\
     &             & 07 Aug 11 &             & 10.2  $\pm$ 0.1       &  ~          &       \\
     &             & 31 Aug 11 &             & 10.2  $\pm$ 0.1       &  ~          &       \\
     &             & 01 Oct 11 &             & 10.1  $\pm$ 0.2       &  ~          &       \\
     &             &           &             &                       &             &       \\
132  & F18203+6210 & 03 Aug 07 &  M7 III     & 11.24 $\pm$ 0.06~$^1$ &  9.9$-$11.0 & 2     \\
     &             & 11 Jul 11 &             & 11.15 $\pm$ 0.08      &  ~          &       \\
     &             & 02 Aug 11 &             & 11.2  $\pm$ 0.1       &  ~          &       \\
     &             & 31 Aug 11 &             & 10.95 $\pm$ 0.07      &  ~          &       \\
     &             & 03 Oct 11 &             & 11.26 $\pm$ 0.08      &  ~          &       \\
     &             &           &             &                       &             &       \\
133  & F18230+6418 & 11 Jul 07 &  M8 III     & 14.58 $\pm$ 0.02~$^1$ & 10.0$-$13.6 & 1  Catalina   \\
     &             & 11 Jul 11 &             & 13.93 $\pm$ 0.04      &  ~          &       \\
     &             & 02 Aug 11 &             & 14.31 $\pm$ 0.03      &  ~          &       \\
     &             & 06 Sep 11 &             & 14.55 $\pm$ 0.03      &  ~          &       \\
     &             & 08 Oct 11 &             & 14.15 $\pm$ 0.06      &  ~          &       \\
     &             &           &             &                       &             &       \\
136  & F18295+6135 & 03 Aug 07 &  M5 III     &  9.9  $\pm$ 0.1~ $^1$ &  9.5$-$10.1 & 2     \\
     &             & 12 Jul 11 &             & 10.10 $\pm$ 0.05      &  ~          &       \\
     &             & 04 Aug 11 &             &  9.93 $\pm$ 0.06      &  ~          &       \\
     &             & 31 Aug 11 &             & 10.08 $\pm$ 0.08      &  ~          &       \\
     &             & 06 Sep 11 &             & 10.15 $\pm$ 0.05      &  ~          &       \\
     &             & 03 Oct 11 &             & 10.00 $\pm$ 0.07      &  ~          &       \\
     &             &           &             &                       &             &       \\
137  & F18307+6153 & 04 Aug 11 &  M0 I       &  8.8  $\pm$ 0.1       &  9.2$-$9.3  & 3     \\
     &             & 31 Aug 11 &             &  8.9  $\pm$ 0.1       &  ~          &       \\
     &             & 03 Oct 11 &             &  8.8  $\pm$ 0.1       &  ~          &       \\
     &             &           &             &                       &             &       \\
138  & F03502+6925 & 16 Dec 13 &  M7  III         &  8.86 $\pm$ 0.06      &             & 2    \\
     &             &           &             &                       &             &       \\
142  & F07550+722  & 17 Dec 13 &  M5  III         &  9.70 $\pm$ 0.06~ $^1$ &             & 2     \\
     &             &           &             &                       &             &       \\
143  & F07554+7246 & 17 Dec 13 &  M2  III         &  8.7  $\pm$ 0.1~ $^1$   &    & 3     \\
     &             &           &             &                       &             &       \\
145  & F08149+6221 & 16 Dec 13 &  M5 III          &  8.6  $\pm$ 0.1~ $^1$    &      & no data  \\
     &             & 16 Dec 13 &             &  8.60 $\pm$ 0.08      &             &       \\
     &             &           &             &                       &             &       \\
\bottomrule
\end{tabular}
\end{table*}

%%%%%%%%%%%%%%%%%%%%%%%

\setcounter{table}{0}
\begin{table*}\footnotesize\centering
 \setlength{\tabnotewidth}{0.5\columnwidth}
 \tablecols{7}
\caption{Log of observations, spectral classification,     magnitudes and variability   class.  C\lowercase{ontinue}  }
\label{tab:tab1}
%\begin{tabular}{@{}lccllcl@{}}
\setlength{\tabnotewidth}{0.95\linewidth}
\begin{tabular}{lccllcl}
\toprule
BIS  & $IRAS$ FSC    &    Date   &  Sp. type   &  $R$ mag~             & $R_{r}$     & Var. class \\
\hline
     &             &           &             &                       &             &       \\
     154 & F12234+6915 & 20 May 1 4 &  M7 III     & 10.04   $\pm$ 0.06           &             & 2     \\
     &             &           &             &                       &             &       \\
155& F13549+7012  & 28 Mar 15 &   M1 III     & 9.2 $\pm$  0.09  &   8.9 &   3 \\
     &             &           &             &                       &             &       \\
156& F15521+7138  & 20 June 15 &   M5 III     & 10.4 $\pm$  0.08  &       10.2      &   2    \\
     &             &           &            &                       &             &       \\
167  & F17278+6416 & 02 Aug 07 & M6 III     & 10.76 $\pm$ 0.08~$^1$ & 10.2$-$10.6 & 2     \\
     &             & 10 Jul 11 &            & 11.1  $\pm$ 0.2       &  ~          &       \\
     &             & 05 Aug 11 &            & 11.2  $\pm$ 0.2       &  ~          &       \\
     &             & 22 Aug 11 &            & 10.8  $\pm$ 0.1       &  ~          &       \\
     &             & 01 Oct 11 &            & 10.9  $\pm$ 0.2       &  ~          &       \\
     &             &           &            &                       &             &       \\
168  & F17313+7033 & 12 Jul 07 & M6 III     &  9.87 $\pm$ 0.07~$^1$ &  9.4$-$9.7  & 2/1     \\
     &             & 08 Jul 11 &            &  9.89 $\pm$ 0.06      &  ~          &       \\
     &             & 07 Aug 11 &            &  9.90 $\pm$ 0.07      &  ~          &       \\
     &             & 16 Aug 11 &            &  9.78 $\pm$ 0.06      &  ~          &       \\
     &             & 09 Sep 11 &            &  9.55 $\pm$ 0.06      &  ~          &       \\
     &             & 16 Nov 11 &            &  9.90 $\pm$ 0.07      &  ~          &       \\
     &             &           &            &                       &             &       \\
170  & F17567+6956 & 01 Aug 07 & M6 III     & 11.15 $\pm$ 0.08~$^1$ & 10.5$-$10.9 & 2  Catalina   \\
     &             & 08 Jul 11 &            & 11.0  $\pm$ 0.1       &  ~          &       \\
     &             & 06 Aug 11 &            & 11.05 $\pm$ 0.07      &  ~          &       \\
     &             & 28 Aug 11 &            & 10.80 $\pm$ 0.07      &  ~          &       \\
     &             & 03 Oct 11 &            & 10.93 $\pm$ 0.08      &  ~          &       \\
     &             & 16 Nov 11 &            & 10.9  $\pm$ 0.2       &  ~          &       \\
     &             &           &            &                       &             &       \\
172  & F05116+6508 & 16 Dec 13 & M4  III  & 10.15 $\pm$ 0.08      &             & 2/3    \\
     &             &           &            &                       &             &       \\
173  & F05125+6652 & 16 Dec 13 & M6  III   & 12.18 $\pm$ 0.05      &     & 2 Catalina  2/3 \\
     &             &           &            &                       &             &       \\
174  & F05141+6509 & 16 Dec 13 & M4 III          &  9.88 $\pm$ 0.06      &             & 2/3   \\
     &             &           &            &                       &             &       \\
184  & F06012+6733 & 20 Jan 10 &N + dust   & 10.4  $\pm$ 0.1~ $^1$ &  ~          & 2  Catalina    \\
  & & 03 Mar 16 &   & 10.28  $\pm$ 0.06~ $^1$ &  ~          &     \\
  & & 14 Feb 17 &   & 9.35  $\pm$ 0.04~ $^1$ &  ~          &    \\
     &             &           &            &                       &             &       \\
%194 & F07003+6815 & 11 Feb 07 & N          &                       &  ~          &       \\
194  & F07003+6815 & 22 Jan 08 & N          &  9.3  $\pm$ 0.2~ $^1$ &  9.4$-$9.6  & 3     \\
 & & 18 Mar 16 &           &  9.45  $\pm$ 0.09~ $^1$ &    &     \\
     &             &           &            &                       &             &       \\
%196 & F07497+6526 & 11 Feb 07 & M8 III     &                       &             & ~     \\
196  & F07497+6526 & 22 Jan 08 & M8 III     & 13.40 $\pm$ 0.04~$^1$ & 11.0$-$12.6 & 1 Catalina    \\
     &             & 07 Dec 08 &            & 13.60 $\pm$ 0.03~$^2$ &  ~          &       \\
     &             &           &            &                       &             &       \\
197  & F08034+6612 & 18 Dec 13 & M5 III          & 10.14$\pm$ 0.05~ $^1$ &             & 3     \\
     &             &           &            &                       &             &       \\
%198 & F08487+6819 & 11 Feb 07 & M5 III     &                       &             & ~     \\
198  & F08487+6819 & 22 Jan 08 & M5 III     & 11.12 $\pm$ 0.03~$^1$ & 10.3$-$10.9 & 2   Catalina   \\
     &             &           &            &                       &             &       \\
\bottomrule
\end{tabular}
\end{table*}

%%%%%%%%%%%%%%%%%%%%
\setcounter{table}{0}
\begin{table*}\footnotesize\centering
 \setlength{\tabnotewidth}{0.5\columnwidth}
 \tablecols{7}
\caption{Log of observations, spectral classification,     magnitudes and variability   class.  C\lowercase{ontinue} }
\label{tab:tab1}
%\begin{tabular}{@{}lccllcl@{}}
\setlength{\tabnotewidth}{0.95\linewidth}
\begin{tabular}{lccllcl}
\toprule
BIS  & $IRAS$ FSC    &    Date   & Sp. type   &  $R$ mag~             & $R_{r}$     & Var. class \\
\hline
     &             &           &            &                       &             &       \\
%199 & F08520+6724 & 11 Feb 07 & M1 III     &                       &             & ~     \\
199  & F08520+6724 & 22 Jan 08 & M1 III     &  8.64 $\pm$ 0.05~$^1$ &  8.8$-$9.1  & 3  Catalina Var.  \\
     &             & 08 Jul 11 &            &  8.6  $\pm$ 0.2       &  ~          &       \\
     &             &           &            &                       &             &       \\
200  & F10093+6847 & 21 May 14 & M5 III     & 10.32    $\pm$ 0.08~   &             & 2     \\
     &             &           &            &                       &             &       \\
201  & F10176+6812 & 20 May 14 & M5 III     & 10.4        $\pm$ 0.1         &       & 2/3   \\
     &             &           &            &                       &             &       \\
203  & F13576+6705 & 21 May 14 & late\,K--M0  III  &  9.2  $\pm$ 0.1~   &   & 3     \\
     &             &           &            &                       &             &       \\
207  & F16148+6532 & 03 Aug 07 & M5 III     & 11.68 $\pm$ 0.04~$^1$ & 11.0$-$12.2 & 2e Catalina     \\
     &             & 06 Jul 11 &            & 11.32 $\pm$ 0.05      &  ~          &       \\
     &             & 07 Aug 11 &            & 11.11 $\pm$ 0.05      &  ~          &       \\
     &             & 22 Aug 11 &            & 11.19 $\pm$ 0.05      &  ~          &       \\
     &             & 11 Oct 11 &            & 11.89 $\pm$ 0.05      &  ~          &       \\
     &             & 22 Oct 11 &            & 11.85 $\pm$ 0.06      &  ~          &       \\
209  & F16317+6603 & 04 Aug 07 &  M5 III     & 10.14 $\pm$ 0.04~$^1$ &  9.7$-$10.4 & 2     \\
     &             & 07 Jul 08 &             & 10.30 $\pm$ 0.05~$^1$ &  ~          &       \\
     &             & 06 Jul 11 &             & 10.41 $\pm$ 0.05      &  ~          &       \\
     &             & 07 Aug 11 &             & 10.33 $\pm$ 0.06      &  ~          &       \\
     &             & 22 Aug 11 &             & 10.42 $\pm$ 0.05      &  ~          &       \\
     &             & 11 Oct 11 &             & 10.35 $\pm$ 0.05      &  ~          &       \\
     &             & 22 Oct 11 &             & 10.25 $\pm$ 0.05      &  ~          &       \\
     &             &           &             &                       &             &       \\
210  & F16499+6532 & 06 Jul 11 &  M2.5 III   &  9.6  $\pm$ 0.1       &  9.4$-$9.6  & 3     \\
     &             & 05 Aug 11 &             &  9.6  $\pm$ 0.1       &  ~          &       \\
     &             & 16 Aug 11 &             &  9.64 $\pm$ 0.07      &  ~          &       \\
     &             & 09 Oct 11 &             &  9.60 $\pm$ 0.5       &  ~          &       \\
     &             &           &             &                       &             &       \\
211  & F17547+6849 & 04 Aug 07 &  M3.5 III   &  9.5  $\pm$ 0.1~ $^1$ & 10.0$-$10.4 & 2     \\
     &             & 07 Dec 08 &             &  9.10$\pm$ 0.06~ $^2$    &       &       \\
     &             & 09 Jul 11 &             & 10.0  $\pm$ 0.1       &  ~          &       \\
     &             & 06 Aug 11 &             & 10.3  $\pm$ 0.1       &  ~          &       \\
     &             & 28 Aug 11 &             & 10.1  $\pm$ 0.1       &  ~          &       \\
     &             & 08 Oct 11 &             &  9.9  $\pm$ 0.1       &  ~          &       \\
     &             &           &             &                       &             &       \\
212  & F17599+6843 & 02 Aug 07 &  M6 III     & 10.04 $\pm$ 0.07~$^1$ &  9.7$-$10.2 & 2     \\
     &             & 09 Jul 11 &             & 10.3  $\pm$ 0.1       &  ~          &       \\
     &             & 07 Aug 11 &             & 10.30 $\pm$ 0.07      &  ~          &       \\
     &             & 28 Aug 11 &             & 10.45 $\pm$ 0.08      &  ~          &       \\
     &             & 08 Oct 11 &             & 10.35 $\pm$ 0.1       &  ~          &       \\
     &             &           &             &                       &             &       \\
213  & F03314+7529 & 18 Dec 13 &  M7 III     & 10.38   $\pm$ 0.06~ $^1$               &             & 2     \\
     &             &           &             &                       &             &       \\
214  & F03315+7340 & 18 Dec 13 &  M5  III   & 13.87   $\pm$ 0.07~ $^1$     &      & 2/1    \\
     &             &           &             &                       &             &       \\
\bottomrule
\end{tabular}
\end{table*}
 
 %%%%%%%%%%%%%%%%%%%%
\setcounter{table}{0}
\begin{table*}\footnotesize\centering
 \setlength{\tabnotewidth}{0.5\columnwidth}
 \tablecols{7}
\caption{Log of observations, spectral classification,     magnitudes and variability   class.  C\lowercase{ontinue} }
\label{tab:tab1}
%\begin{tabular}{@{}lccllcl@{}}
\setlength{\tabnotewidth}{0.95\linewidth}
\begin{tabular}{lccllcl}
\toprule
BIS  & $IRAS$ FSC    &    Date   & Sp. type   &  $R$ mag~             & $R_{r}$     & Var. class \\
\hline
     &             &           &            &                       &             &       \\
     215 &  F03328+7717 & 15 Jan 16 &  M3 III  & 12.54   $\pm$ 0.04~ $^1$     &    & 3    \\
     &             &           &             &                       &             &       \\
216 & F03353+8722 & 17 Dec 13 &  M9/S   & 15.3   $\pm$ 0.1~ $^1$               &             & 2     \\
     &             &           &             &                       &             &       \\
219 & F04148+7504 & 17 Dec 13 &  M4--M5 III       &  10.92      $\pm$ 0.06~ $^1$   &    & 2e     \\
     &             &           &             &                       &             &       \\
222  & F04210+7517 & 19 Jan 10 &  N + dust   & 18.5  $\pm$ 0.2~$^1$ &  ~          & 2     \\
     &             & 16 Dec 13 &             & 20.6 $\pm$ 0.3~$^1$     &             &   \\
     &             & 13 Jan 16 &             & 20.9 $\pm$ 0.2~$^1$     &             &   \\
     &             & 15 Feb 17 &             & 20.5 $\pm$ 0.1~$^1$     &             &   \\
     &             & 15 Feb 17&             & 17.60 $\pm$ 0.06~$^1$     &             &  I mag  \\
     &             &           &             &                       &             &       \\
224  & F04423+7314 & 17 Dec 13 &  M5  III         &11.01$\pm$ 0.05~ $^1$  &             & 2/3   \\
     &             &           &             &                       &             &       \\
226 & F04464+7900 & 17 Dec 13  & M7 III          & 12.83~$^1$                 &             & 2 \\
     &             &           &            &                       &             &       \\
228  & F04488+7831 & 18 Dec 13 & M1 III          & 8.9 $\pm$ 0.1~ $^1$   &             & 3     \\                  
     &             &           &            &                       &             &       \\
247  & F06561+7359 & 18 Dec 13 & M4.5(E)    &  9.71  $\pm$ 0.1~ $^1$             &             & 3     \\
     &             &           &            &                       &             &       \\
248  & F06563+7354 & 18 Dec 13 & M4 III          &  9.00  $\pm$ 0.1~ $^1$    &    & 3     \\
     &             &           &            &                       &             &       \\
%255 & F09311+7844 & 22 Apr 08 & M3 III     &                       &             & ~     \\
255  & F09311+7844 & 08 Jul 11 & M3 III     &  9.20 $\pm$ 0.05      &  9.3$-$9.6  & 2/3     \\
     &             & 06 Aug 11 &            &  9.25 $\pm$ 0.07      &  ~          &       \\
     &             & 13 Aug 11 &            &  9.2  $\pm$ 0.1       &  ~          &       \\
256  & F10289+7815 & 22 Apr 08 & M3 III     &                       &             & 3     \\
     &             & 06 Aug 11 &            &  8.7  $\pm$ 0.2       &  ~          &       \\
     &             & 13 Aug 11 &            &  8.9  $\pm$ 0.1       &  ~          &       \\
     &             & 17 Mar 12 &            &  7.54 $\pm$ 0.08 I ~ $^1$   &             &       \\
     &             & 17 Mar 12 &            &  8.56 $\pm$ 0.08 R ~ $^1$   &             &       \\
     &             & 17 Mar 12 &            &  9.35 $\pm$ 0.07 V ~ $^1$   &             &       \\
     &             & 17 Mar 12 &            & 11.2 $\pm$ 0.1 B ~ $^1$   &             &       \\
     &             &           &            &                       &             &       \\
258  & F11594+7309 & 20 May 14 & M0 III     &  8.64 $\pm$ 0.1    &             & 3  \\
     &             &           &            &                       &             &       \\
260  & F12198+7909 & 08 Jul 11 & M1 III     &  8.9  $\pm$ 0.1       &  8.9$-$9.0  & 3     \\
     &             &           &            &                       &             &       \\
264  & F13410+8000 & 22 Apr 08 & M4  & 11.82 $\pm$ 0.04~$^1$ & 11.6$-$12.4 & 2e    \\
     &             & 07 Jul 11 &            & 11.51 $\pm$ 0.07      &  ~          &       \\
     &             & 04 Aug 11 &            & 11.97 $\pm$ 0.06      &  ~          &       \\
     &             & 11 Aug 11 &            & 11.98 $\pm$ 0.06      &  ~          &       \\
     &             & 09 Oct 11 &            & 11.35 $\pm$ 0.05      &  ~          &       \\
     &             &           &            &                       &             &       \\
267  & F14366+8058 & 20 May 14 & M4  III         &  9.8  $\pm$ 0.1     &             & 1 \\
     &             &           &            &                       &             &       \\
\bottomrule
\end{tabular}
\end{table*}

%%%%%%%%%%%%%%%%%%%%%%

\setcounter{table}{0}
\begin{table*}\footnotesize\centering
 \setlength{\tabnotewidth}{0.5\columnwidth}
 \tablecols{7}
\caption{Log of observations, spectral classification,     magnitudes and variability   class.  C\lowercase{ontinue} }
\label{tab:tab1}
%\begin{tabular}{@{}lccllcl@{}}
\setlength{\tabnotewidth}{0.95\linewidth}
\begin{tabular}{lccllcl}
\toprule
BIS  & $IRAS$ FSC    &    Date   & Sp. type   &  $R$ mag~             & $R_{r}$     & Var. class \\
\hline
     &             &           &            &                       &             &       \\
271  & F15152+7303 & 20 June 15 & M4 III         &  10.5$\pm$ 0.2     &   10.8    & 2 \\
     &             &           &            &                       &             &       \\
275  & F16549+7935 & 03 Aug 07 & M3 II      &  9.08 $\pm$ 0.07~$^1$ & 8.7$-$8.9   & 2/3    \\
     &             & 06 Jul 11 &            &  8.93 $\pm$ 0.07      &  ~          &       \\
     &             & 05 Aug 11 &            &  9.0  $\pm$ 0.1       &  ~          &       \\
     &             & 24 Aug 11 &            &  8.9  $\pm$ 0.1       &  ~          &       \\
     &             & 27 Nov 11 &            &  9.1  $\pm$ 0.1       &  ~          &       \\
     &             &           &            &                       &             &       \\
276  & F16583+7852 & 03 Aug 07 & M3.5 III   & 10.62 $\pm$ 0.05~$^1$ & 10.4$-$11.4 & 2     \\
     &             & 07 Jul 08 &            & 10.60 $\pm$ 0.05~$^1$ &  ~          &       \\
     &             & 07 Jul 11 &            & 10.73 $\pm$ 0.05      &  ~          &       \\
     &             & 06 Aug 11 &            & 10.64 $\pm$ 0.05      &  ~          &       \\
     &             & 24 Aug 11 &            & 10.70 $\pm$ 0.05      &  ~          &       \\
     &             & 05 Oct 11 &            & 10.65 $\pm$ 0.05      &  ~          &       \\
     &             &           &            &                       &             &       \\
285  & F18038+8302 & 04 Aug 07 & M3 III     &  9.90 $\pm$ 0.03~$^1$ &  9.8$-$10.0 & 2/3     \\
     &             & 09 Jul 11 &            &  9.90 $\pm$ 0.05      &  ~          &       \\
     &             & 04 Aug 11 &            &  9.88 $\pm$ 0.05      &  ~          &       \\
     &             & 24 Aug 11 &            &  9.90 $\pm$ 0.04      &  ~          &       \\
     &             & 05 Oct 11 &            &  9.92 $\pm$ 0.03      &  ~          &       \\
     &             & 26 Nov 11 &            &  9.69 $\pm$ 0.05      &  ~          &       \\
     &             &           &            &                       &             &       \\
\bottomrule
\end{tabular}
\end{table*}

%%%%%%%%%%%%%%%%%%%%%%
\end{document}